\documentclass[twocolumn,preprintnumbers,aps,superscriptaddress,showpacs,nofootinbib]{revtex4-1}

\def\bea{\begin{eqnarray}}
\def\eea{\end{eqnarray}}
\def\ben{\begin{equation}}
\def\een{\end{equation}}
\def\benu{\begin{enumerate}}
\def\enu{\end{enumerate}}
\def\bei{\begin{itemize}}
\def\eei{\end{itemize}}
\def\blue#1{{\em #1}}

\def\n{\rho}

\def\sss{\scriptscriptstyle\rm}

\def\1var{(\bx_1...\bx\N)}

\def\br{{\bf r}}

\def\bx{{x}}

\def\s{_{\sss S}}
\def\xc{_{\sss XC}}

\def\N{_{\sss N}}

\def\sph_int{ {\int d^3 r}}

\usepackage{graphicx}
\usepackage{psfrag}
\usepackage{amsmath}
\usepackage{bm}

\usepackage{epsfig}

\input epsf

\def\sec#1{\section{#1}}

\begin{document}
\title{Perspective on density functional theory}
\date{\today}
\author{Kieron Burke}
\affiliation{Department of Chemistry, 1102 Natural Sciences 2, UC Irvine, CA 92697, USA}
\begin{abstract}
Density functional theory (DFT) is an incredible success story.  
The low computational cost, combined with useful (but not yet chemical) accuracy,
has made DFT a standard technique in most branches of chemistry and materials
science.
Electronic
structure problems in a dazzling variety of fields are currently being tackled.
However, DFT has many limitations in its present form: Too many approximations,
failures for strongly correlated systems, too slow for liquids, etc.
This perspective reviews some recent progress and ongoing challenges.
\end{abstract}

\maketitle

\begin{figure}[htb]
\begin{center}
\includegraphics[width=3.5in]{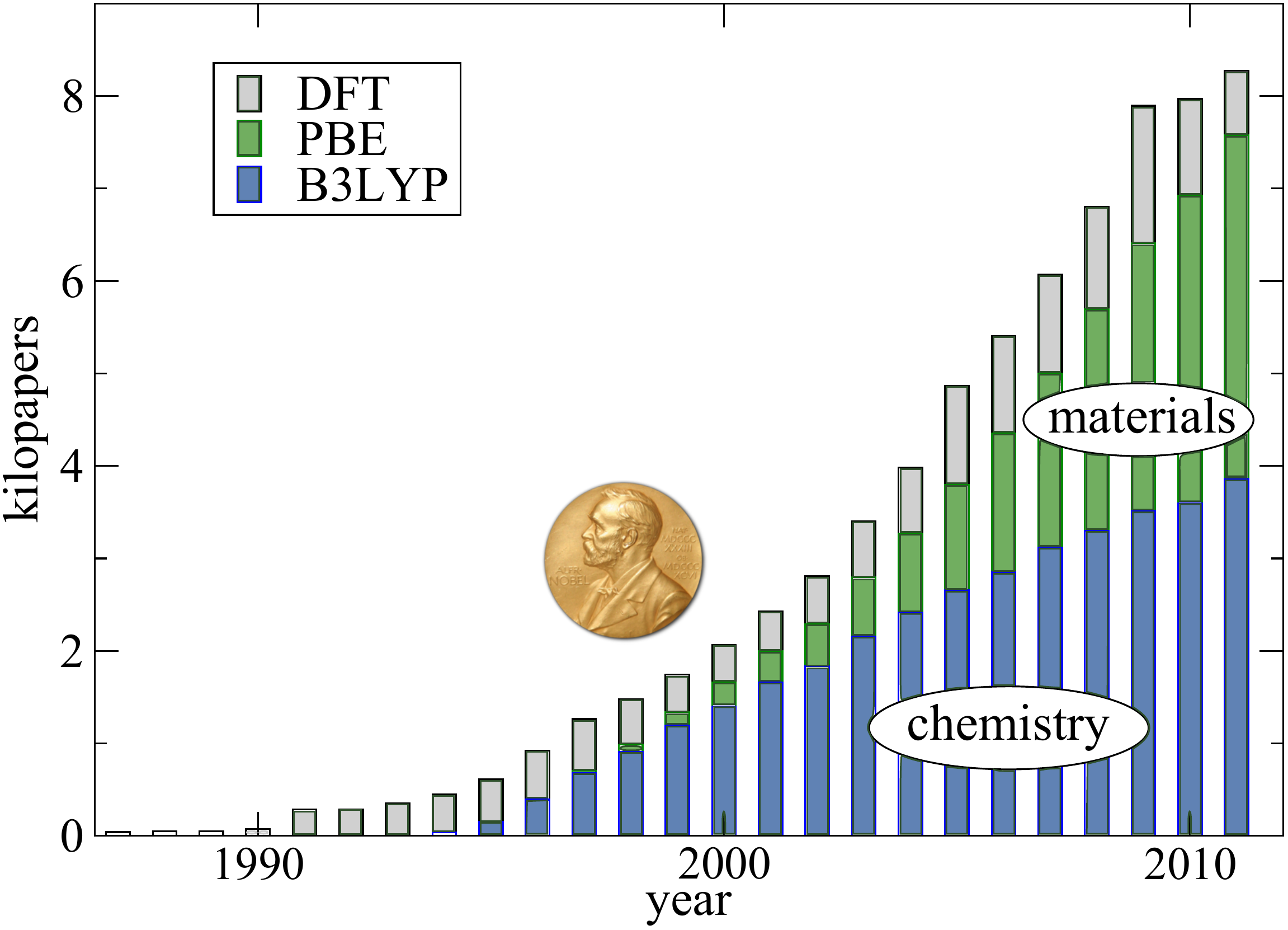}
\caption{Numbers of papers when DFT is searched as a topic in Web of Knowledge
(grey), B3LYP citations (blue), and PBE citations (green, on top of blue).}
\label{papers}
\end{center}
\end{figure}

\sec{Introduction}
Over the past 20 years, DFT has become a much used tool in most branches of chemistry.
Many experimental investigations in organic and inorganic chemistry routinely
include such calculations, using a popular code, a standard basis, and 
a standard functional approximation \cite{CZFF11}. 
A similar transformation is now underway in materials
science where, in the past decade, improvements in both hardware and codes
have made it possible to perform systematic comparisons with experiment across
large ranges of materials, learning which approximations work and why,
and allowing for true first-principles predictions of properties.
Among notable recent successes are the prediction of new catalysts \cite{VBGG12}
and new Li battery materials \cite{MOC11}
in the Materials Genome Project.
A complementary aspect of this story is shown in Fig. \ref{papers}, which
plots the number of papers given by Web of Knowledge when DFT is searched as a topic (grey bars).
This will soon reach 10,000 per year, vindicating the 1998 Nobel prize in chemisty, which went
to Walter Kohn \cite{K99}
for inventing the theory and to John Pople \cite{P99} for making it
accessible through popular computational packages.
The figure 
also marks the fraction of papers citing B3LYP \cite{B93,B88,LYP88}, currently the most popular
approximation in chemistry,  and PBE \cite{PBE96}, the most popular approximation in materials.
Clearly, applications to materials
will soon outstrip those in chemistry.

This perspective is for a general audience, and
focuses on fundamental general aspects of
DFT, rather than detailed computational procedures and results for specific systems.
Because DFT is now applied so broadly, no such article can hope to be comprehensive.
The topics covered here are designed to give a flavor of how the field works, and 
are mostly those I have personally worked in.
Great longer reviews exist  for experts \cite{FNM03,DG90,PY89} and for users\cite{KH03},
as well as introductions for neophytes \cite{ABC,BW11}.
An excellent complementary perspective, including computational issues and non-DFT approaches,
is Ref. \cite{S10}.
I work always within the Born-Oppenheimer approximation, the non-relativistic limit, and discuss 
only the basis-set limit of calculations.  I use atomic units (lengths in
Bohr, energies in Hartree) except where otherwise specified.
For simplicity, I refer always to density functionals, but modern calculations use
spin-density functional theory, a generalization \cite{BH72}.

\sec{A brief history}
\label{hist}

Our story begins in 1926 with the creation
of Thomas-Fermi theory \cite{T27,F28}, an approximate method for
finding the electronic structure of atoms using just the one-electron
ground-state density, $\n(\br)$,  but too crude
to bind molecules \cite{T57}.  In the 50's, Slater \cite{S51}
intuitively combined this idea with Hartree's orbital method \cite{H28} in the X$\alpha$
scheme.  Later, the Hohenberg-Kohn (HK) theorem \cite{HK64} proved that an exact method
based on $\n(\br)$
exists in principle. The modern version in use today is Kohn-Sham (KS) DFT,
which defines self-consistent equations that must be solved for 
a set of orbitals whose density, $\n(\br)$ is defined to be exactly that of the real
system\cite{KS65}.  In these equations, a small but vital contribution to the
energy, the exchange-correlation (XC) energy, must be given 
in terms of $\n(\br)$.  In principle, and for small systems,
this functional can be found exactly, but turns out to be more
expensive than direct solution of the Schr\"odinger equation \cite{SWWB11}. 
In practical calculations, the XC contribution is approximated,
and the results are only as good as the approximation used.

The simplest XC approximation is 
the local density approximation
(LDA) \cite{KS65} which became the popular standard in calculations on solids in the 70's and 80's.
But molecules in LDA are
typically overbound by about 1 eV/bond, and in the late
1980's generalized
gradient approximations (GGA's) \cite{P86} produced an accuracy that was useful in
chemical calculations.   In the early 90's, hybrids were introduced
by Becke \cite{B93}, replacing a fraction of GGA exchange\cite{B88} with Hartree-Fock exchange,
leading to the ubiquitous B3LYP \cite{LYP88}, the most popular approximation in
use in chemistry today. On the other hand, the PBE GGA \cite{PBE96} has come to dominate
applications to extended systems (materials).
We denote these three (LDA, PBE, and B3LYP) as the {\em standard} approximations,
meaning that they are the most popular examples of each type of approximation,
and dominate the user market, as shown by Fig. \ref{papers}.

To give an idea of how much (or how little) progress is made in DFT development,
we go back to the year of the Nobel prize, 1998.
Some of the most prominent practical difficulties from back then
include
(i) gaps of bulk solids are underestimated, 
(ii)
van der Waals missing from popular functionals,
(iii)
strongly correlated systems poorly treated,
and (iv) no good scheme for excitations.
Of course, there are many others, but this sample will give us some idea of how things work.
The rest of this article interweaves sections labeled {\em progress}, indicating
areas where substantial progress has been made since that time, with {\em general
background}, explaining relevant concepts,
and {\em challenges}, areas that need improvement and where we can hope for progress in the next decade.

\sec{Progress: materials and nanoscience}
\label{nano}

As shown in Fig. \ref{papers}, materials applications now share the limelight
with chemistry.  Nanoscience completely interweaves these two, such as
when a molecule is adsorbed on a surface.  Many areas of materials
research, especially those related
to energy, desperately need input from electronic structure methods.
Most calculations of materials use codes that differ from traditional quantum
chemistry codes, because they employ plane waves that satisfy periodic boundary conditions \cite{M08}.
Such calculations
converge much faster to the bulk limit than by taking ever larger finite clusters of
the material.  In fact, almost all popular codes are designed either for finite molecular
systems \cite{ABHH89} or for extended bulk systems \cite{QE2009},
although codes that treat both are beginning to
appear \cite{FHIaims,MHJ05}. 

In solid-state physics, DFT has always been more popular than traditional direct
solution of the Schr\"odinger equation,
because the Hartree-Fock approximation has unpleasant
singularities for zero-gap materials, i.e., metals.
The next logical step \cite{P01} beyond the standard approximations are
meta-GGA's, which include the kinetic energy density as an input, and can yield
accurate ground-state energies simultaneously for molecules, solids, and surfaces \cite{SMRK11}.
But an outstanding
failure has been DFT's inability to provide good estimates of the fundamental gaps
of semiconductors and insulators, a crucial quantity for much
materials research, such as impurity levels in doped semiconductors \cite{GSSG11}. 

The fundamental (or transport) gap is $I-A$, where $I$ is the ionization potential
and $A$ the electron affinity.  It is well-established that the KS gap, i.e., the 
difference between the KS HOMO and LUMO energies, does {\em not} equal this value,
even with the exact XC functional \cite{PPLB82,Pb85}.  Calculations suggest it is typically 
substantially smaller (by about 50\%) \cite{GMR06}. 
The LDA and GGA approximations yield accurate
values for the KS gap.  But if $I-A$ were calculated by adding and subtracting electrons
from a large cluster of the material, it would match the KS gap within such approximations,
because they incorrectly allow electrons
to completely delocalize over insulating solids.
Such approximations lack a derivative discontinuity that
accounts for this difference \cite{Pb85}.

A practical way around this problem is to use a hybrid, but treating the exchange 
term as orbital dependent, via generalized KS theory \cite{SGVM96}.  With mixing parameters
of about 20\%, this typically yields much better gaps for semiconductor materials.
A recent functional, called HSE \cite{HSE03,HS04}, not only mixes in some HF, but also performs a 
range-separation.  Based on exact theorems of Savin\cite{S96}, the short-range
part of the HF is treated exactly, while the long-range contribution is treated
by approximate DFT.  The resulting functional, used to calculate gaps in the
generalized scheme, appears to work accurately
for a large variety of moderate-gap semiconductors \cite{HS04}, overcoming the problems
of LDA and GGA.  It yields accurate fundamental gaps when excitonic effects are negligible,
and is closer to optical gaps when they are not
\cite{BINB08,JCL11}.

This kind of progress is very welcome in an era in which we must accurately tackle
moderately correlated systems.  Transition metal oxides play vital roles in many
energy-related problems, such as creation of efficient photovoltaics \cite{SLYC06}.  The standard
approximations over-delocalize the $d$-electrons, leading to highly incorrect
descriptions.  Many practical schemes (HSE \cite{HSE03},
GGA+U \cite{AZA91,KCSM06}, Dynamical Mean-Field Theory (DMFT) \cite{GKKR96}) can correct
these difficulties, but none has yet become a universal tool of known performance
for such systems.  Very recently, a promising non-empirical scheme has been suggested
for extracting gaps using {\em any} approximate functional \cite{ZCMH11}.

An alternative approach to directly tackling such problems is to study them in
simpler situations, and test suggested remedies on cases where exact, unambiguous
answers can be easily obtained.  Recently, an extremely powerful technique for direct
solution of many-body problems, called DMRG, has been adapted to tackle a one-dimensional
analog of the real world
\cite{WSWB11}.  DMRG is powerful enough to calculate the exact
XC functional on systems of 100 atoms or more\cite{SWWB11}.  We will see how the exact
functional deals with strongly correlated insulators and which new approximations
are working for the right reasons.

For more than a decade, researchers have been performing DFT calculations of
molecular conductance, calculating the current in response to a bias applied
to a molecule caught between two metal leads \cite{VPL00}. 
This problem is prototypically
difficult for present electronic structure methods \cite{KCBC08}. 
This is a steady state situation, not a ground-state one. Model Hamiltonians
typically used to study this kind of physics are insufficiently accurate, as several hundred
atoms must be treated to achieve chemical realism \cite{NR03}.  
A standard approach, combining the Landauer formula \cite{L57} with ground-state
DFT and often called non-equilibrium Green's functions (NEGF) \cite{TGW01}, is typically used, and
avoids adjustable parameters.  But especially in the case of organic molecules, there
is every reason to believe that, for the reasons given above about bulk gaps, our standard
approximations are simply wildly inaccurate for this problem.  Recent calculations \cite{SZVD05,TFSB05}
including corrections based on non-DFT many-body corrections \cite{QVCL07}, 
and simple exact
results for simple models \cite{BLBS11}, are strong evidence that, at least for weak bias, the
standard approach should yield accurate currents, once better DFT approximations
are used \cite{LBBS11}.

\sec{Background: Exact theory}
\label{exact}


A KS calculation appears very similar to a Hartree-Fock calculation, but 
there's a crucial difference.
The effective potential is {\em defined} to be one which makes the
one-particle density $\n(\br)$ be the {\em exact} density of the
system.  By virture of the HK theorem \cite{HK64,L79}, one can write
\ben
E = T\s + U + V_{nuc} + E\xc[\n]
\een
where $T\s$ is the energy of the KS orbitals, $U$ is the Hartree 
(a.k.a. Coulomb) energy, $V_{nuc}$ is the attraction to the nuclei,
and $E\xc[\n]$ is everything else, defined to make the above exact.
Then one finds the minimizing orbitals are given by the celebrated
KS equations\cite{KS65}:
\ben
\left( -\frac{1}{2}\nabla^{2}+v\s(\br)\right)\phi_{j}(\br) =
\epsilon_{j} \phi_{j}(\br) \ ,
\een
and 
\ben
v\s(\br) = v_{nuc}(\br) + \int d^3r'\, \frac{\n(\br')}{|\br-\br'|}
+ v\xc[\n](\br)
\een
and $v\xc(\br) = \delta E\xc/\delta \n(\br)$.  Thus, if the
XC energy is known as a functional of the density, these form a
closed set of self-consistent equations yielding the exact answer
to the electronic structure problem, without ever calculating 
the electron-electron repulsion directly!

\begin{figure}[htb]
\begin{center}
\includegraphics[width=3.5in]{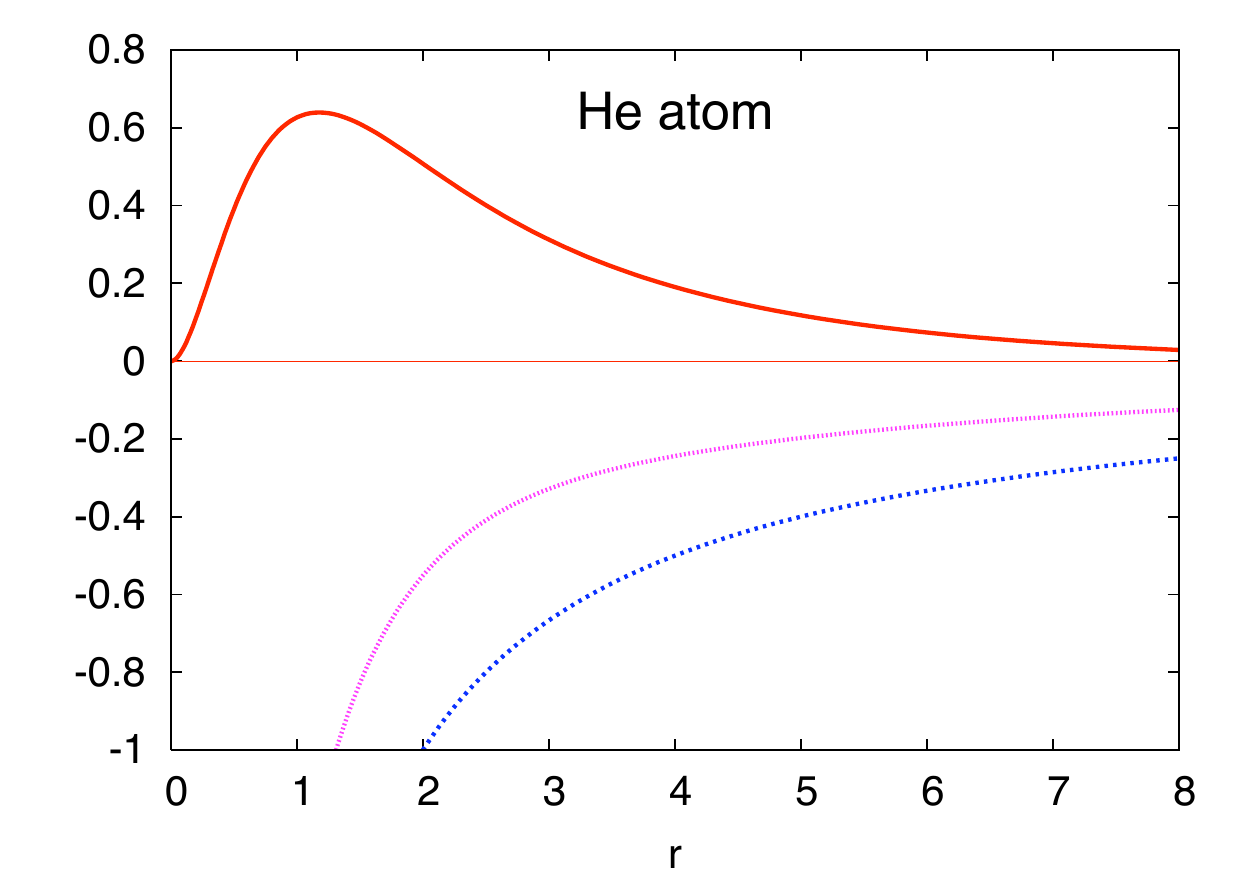}
\caption{Exact radial density (red) and potentials (nuclear
is blue, KS is pink) for the He atom in atomic
units (Bohr radii, 0.529$\AA$.).   Two non-interacting fake electrons,
doubly occupying the 1s orbital of the pink potential, produce the {\em exact}
ground-state density shown above. I thank Cyrus Umrigar for this density\cite{UG94}.}
\label{vs}
\end{center}
\end{figure}
In Fig. \ref{vs}, I show the {\em exact} KS potential for a He atom.  The blue
line in the bottom half is $v_{nuc}(r)=-2/r$ in atomic units, the attraction
of the electrons to the nucleus.  In the top half, the red line indicates the
exact radial density of the He atom, found by using sophisticated wavefunction
techniques to directly solve the Schr\"odinger equation \cite{FHM84}.  But the pink dashed
line in the bottom is then the unique potential experienced by two fictitious
{\em non-interacting} electrons that makes them have the exact ground-state density,
i.e. $\phi_{1s}(\br) = {\sqrt{\n(\br)/2}}$. 
Every practical KS DFT calculation approximates $v\s(\br)$, and {\em never} calculates
the true many-electron wavefunction. \footnote{If you wish to annoy and confuse a traditional
quantum chemist, ask ``How much correlation is there in the KS wavefunction?"}

Several relevant points about these equations include:

\noindent
$\star\,$ {\em Failures} of DFT are due to failures of approximations.
The {\em exact} $E\xc[\n]$ yields $E$ and $\n(\br)$ exactly. 

\noindent
$\star\,$ Ground-state DFT calculations, even with the exact XC, yield
{\em only} $E$ and $\n(\br)$ and any property that can be extracted from
them (such as the ionization potential).

\noindent
$\star\,$ The KS {\em energies and orbitals} replace those of HF for understanding
chemical reactivity \cite{BG97}, 
even though they appear as mere
constructs for $\n(\br)$.  We also now know that orbital energy differences
approximate {\em optical} transitions, via TDDFT (Sec \ref{TDDFT}).

\noindent
$\star\,$ For most applications, we care only
about $E$ as a function of the nuclear
coordinates.  This determines bond lengths,
angles, vibrational frequencies,
all reaction energies, transition-state barriers, etc. , i.e., most of
the properties of interest about a molecule or a material.

%

\sec{Challenge: Larger systems}
\label{OFDFT}

Although KS-DFT is just about the quickest way to get useful quantum
calculations of electronic structure, its still too slow for a crucial
application: MD simulations of liquids.  
There is tremendous interest in performing calculations that mix
KS-DFT with other methods.  For example, molecular mechanics with classical
force fields can routinely handle millions of atoms, but making and
breaking bonds is usually unreliable.  On the other hand, the field of
ab initio molecular dynamics (AIMD), using the Car-Parrinello method \cite{CP85} of
combining DFT with MD, is thriving, but limited to a few hundred atoms
per simulation.  Thus there is tremendous desire to perform mixed
{\em QM/MM} simulations, in which the chemically active part of a larger
system is treated in DFT.   Many methods have been proposed, perhaps
the most popular being ONIOM \cite{SHFM96}, but many questions remain, especially
when the interface between the classical and DFT regions involve
covalent bonds.  This is a very active area of research.

A related theme is that of {\em orbital-free DFT}, which looks backward to the
original form of a pure DFT, as approximated by TF theory, and shown to be
formally exact by the HK theorem.  In a modern context, this means producing
a sufficiently accurate approximation (much more accurate than TF theory)
for the KS kinetic energy ($T\s$ in the language of
Sec. \ref{exact}), thereby avoiding the need to
solve the KS equations.
If this could be done, then all DFT calculations would run much more
quickly, and AIMD could be performed on much larger systems.
This has been an active area of research for many decades \cite{DG90}, with various
empirical approaches being tried for such approximations.  Most
(if not all) attempts use reasoning similar to that used successfully
to approximate XC contributions, but there is nothing to suggest (and 
much evidence to the contrary) that such methods can produce a general
purpose approximation sufficiently accurate for this purpose.  Possibly
a lower degree of accuracy could be acceptable for the larger system,
and then a traditional KS treatment applied to the chemically active region.
This  {\em embedding} idea has been tried, especially when the system is weakly
bonded to the environment \cite{WW93,GBM11}.

Many of these ideas are unified in a new approach to the issue of
subsystems within electronic structure, called partition DFT (PDFT).
Originally developed by Cohen and Wasserman to resolve difficult issues
in chemical reactivity theory \cite{CW07}, this formalism unifies
several distinct concepts in chemistry, including atoms in molecules, effective
charges, chemical identity, localization of bonds, etc.  To date,
only calculations on model systems\cite{EBCW10}
and diatomic molecules \cite{NWW11}
have appeared,
but interest is growing
rapidly \cite{HC11}
An interesting challenge is to adapt PDFT to include external
electric fields, as it seems ideally suited to the molecular electronics
problems of Sec. \ref{nano}.

\sec{Background: The users' lament}
\label{lament}
But the popular success of DFT has bred its own set of problems.
As codes become faster and easier to use, DFT is applied to a huge number
of situations.  Fig. \ref{papers} includes applications to protein folding \cite{GASM07}, astrophysics \cite{MCKS07},
dyes \cite{PJP07}, and dirt \cite{PKS07}, to name just a few.
As experience builds with each given
functional, the accuracy and reliability comes to be known, as well
as the
qualitative failures.  For example, it was very early recognized that 
standard approximations do not yield long-range dispersion \cite{JG89}.
The local nature of the standard approximations implies an exponential
decay of the interaction.  Progress in this area only came
recently, and is described in Sec. \ref{weak}.
Another example is that of anions, which are
technically unbound with standard approximations\cite{SRZ77}, and yet for which 
accurate results can still be obtained \cite{GS96}.  Only recently has
this puzzle been addressed and a solution
proposed \cite{KSB11}.

Naturally, users would
like an all-purpose tool that provides answers of a prescribed quality
in all situations.  Present DFT calculations are a far cry from this.
Table I shows a list of features that most users will unfortunately
recognize.
Throughout its history, DFT has provided approximations that work for
some problems and fail for others, in largely mysterious ways.   At any
given moment, the most popular approximations fail for the most
interesting systems, such as the moderately correlated oxides of
Sec. \ref{nano}.
\begin{table}
\begin{tabular}{|c|c|c|}
\hline
:(~~&no simple rule for reliability&$\heartsuit$\\
:(~~&no systematic route to improvement&$\heartsuit$\\
:(~~&decades between each generation&$\heartsuit$\\
:(~~&full of arcane insider jargon&$\heartsuit$\\
:(~~&too many approximations to choose from&$\heartsuit$\\
:(~~&can only be learned from a DFT guru&$\heartsuit$\\
\hline
\end{tabular}
\caption{List of things users despise about DFT calculations.  Please rank in
order of induced frustration.  The extreme left column indicates the users response
to these 'features', the right denotes that of developers.}
\label{frust}
\end{table}

\sec{Progress: Excitations}
\label{TDDFT}
Because an excited-state density does not uniquely determine the
potential \cite{GB04}, there is no general analog of HK for excited states.
Many different ways to approach the calculation of
excited states in DFT have developed over the years, including ensembles \cite{GOK88,OGK88} 
$\Delta$SCF \cite{ZRB77}, min-max principles\cite{LN99}, and others.  
However, since the mid-90's, TDDFT has become extremely
popular \cite{BWG05}. Because the methodology is technically very similar to
that of TDHF, TDDFT was very rapidly implemented in quantum chemical codes
such as Turbomole \cite{BA96}, and is now a standard part of any code.
A useful reference and pedagogical tool is \cite{MNRB06}.

TDDFT is based on a similar (but distinct) theorem to HK.  he Runge-Gross
theorem \cite{RG84} establishes that, in a time-dependent quantum problem,
all observables are functionals of the time-dependent density 
(under certain conditions).
 Then, considering the linear-response of
a molecule to a time-dependent electric field, one finds simple formulas
that correct the KS eigenvalue differences into the (in principle, exact)
optical excitations of the system \cite{C96}, but requiring (of course) another
unknown functional, called the XC kernel \cite{GK85}.  All practical calculations
employ the adiabatic approximation, and almost
all use the same approximation
for both the ground-state and the TDDFT calculation \cite{EFB09}.

Practical TDDFT often produces good excitation spectra.  Typical errors
in individual energies are higher than for the ground-state (0.4 eV), but
properties are excellent (bond lengths, dipole moments, etc.), and even
roughly accurate spectra can be sufficient to identify the dominant
excitations in the optical spectra of large molecules \cite{EFB09}.  About 10\% of all
DFT calculations now include TDDFT as well, as its computational cost is
not many times more than a single calculation of a ground-state energy.
The procedure can even be extended to continuum states \cite{FB09}
to accurately predict electron-atom elastic scattering.

But, just as with ground-state DFT, along with success come challenges.
The standard functionals are inaccurate for charge
transfer excitations, leaving them so low so that they contaminate 
other parts of the spectrum.  Double excitations are also excluded 
by the adiabatic approximation.  Calculations of the optical response of solids
look very similar to RPA (see below) if using standard approximations, 
which are insufficiently non-local in the case of a solid.  
For applications beyond linear response (strong fields),
the standard approximations cannot be used
because of their poor quality potentials.  Furthermore, sometimes
the quantity of interest is not a simple functional of the
time-dependent one-electron density, such as the double ionization 
probability \cite{L01}.

But all these
are open areas of research and progress is being made continuously.
Many papers have been published on the charge transfer issue, and
several range-separated schemes seem to handle these excitations
well \cite{SKB09}. 
The basic form of the kernel needed to handle double excitations
was deduced years ago \cite{MZCB04}, 
and has been generalized and systematized
successfully \cite{HIRC11}.
With much effort, the nature of the error for solids
was deduced \cite{ORR02}, 
and a new approximation to the kernel (the nanoquanta
kernel) can be found from many-body perturbation theory \cite{RORO02}. 
In a very recent development, a bootstrap approximation
for the kernel in terms of the dielectric function appears to produce
accurate excitons in all but the largest-gap insulators \cite{SDSG11}.

\sec{Background: First principles or unprincipled?}
\label{first}

Over the decades since the introduction of the Schr\"odinger equation, many 
excellent methods have been developed
for directly solving the electronic structure problem,
including configuration interaction, M\"oller-Plessset perturbation theory,
the coupled-cluster expansion, and quantum Monte Carlo \cite{Levine}.
Such methods rarely suffer from any of the difficulties listed in Table 1.
But because of the coupling between coordinates in the many-electron Schr\"odinger equation,
the computational cost of such methods is usually significantly higher than
that of DFT.  Loosely, the more accurate the method,
the more rapidly the cost rises with number of atoms.  Thus, without some algorithmic 
breakthrough, DFT will always allow more atoms (often by a factor of 10) to be treated,
no matter how fast our processors get, or how many we have.  

This does not make direct solutions obsolete.  
They provide crucial benchmarks for
testing approximate functionals for smaller systems, and give crucial insight into
the nature of errors, both quantitative and qualitative.  
In chemistry, it is traditional to refer to standard approaches as {\em ab inito}, while
DFT is regarded as empirical.  Because solid-state calculations are more
demanding, for many
decades DFT was the only possible approach.  Thus DFT calculations are referred to
as {\em ab initio} in solid-state physics and materials science.
This is why there is a solid-state code ABINIT \cite{GAAB09}
which performs {\em only} DFT calculations.
\footnote{When teaching, I explain that DFT is some acronym
for unreliable, while {\em ab initio} is Latin for {\em too expensive}.}

A sore point is whether or not approximate DFT should be called {\em empirical}.
Even if an approximate functional
includes parameters that have been fit to some data set, once the
final form has been written down, that approximation can be applied to
every possible electronic structure problem, without adjusting parameters
for each specific calculation.   Thus DFT, with a fixed
approximate functional, is still first principles, in the sense that
the user only chooses the atoms, and the computer
predicts all properties of the molecule or solid.  

As mentioned above, the first approximation was LDA \cite{KS65}, and the formula for this is
determined by properties of the uniform electron gas.  No-one disputes
that DFT with LDA is non-empirical \cite{D30,PW92}.  But even just
the next step up Jacob's ladder \cite{P01} of functional sophistication,
the GGA has no unique form.  There are two major schools of thought here.
Purists like to use exact conditions of quantum mechanics to derive the
parameters in their approximate functionals, and so claim to be non-empirical.
This school has been championed by John Perdew, with a lifetime of very
successful approximations \cite{PZ81,PW92,PBE96}.  On the other hand, pragmatists
like Axel Becke and Bob Parr have allowed one or two parameters to be
fit to specific systems, such as in B88 for exchange \cite{B88} and LYP for
correlation \cite{LYP88}.  Such approximations have been based on sound physical
reasoning {\em underlying} the structure of the approximation.  I have
even had the pleasure of {\em deriving} some of these parameters much later \cite{BEP97,EB09}.
By fitting, one usually
finds higher accuracy for systems similar to those fitted (often by a factor of 2),
but greater
inaccuracies far away.  For example, LYP correlation \cite{LYP88} works very well
as part of B3LYP in chemistry, but fails badly for bulk metals.
The PBE approximation \cite{PBE96} works passably well for many materials purposes, but can
be a factor of 2 or more worse than BLYP for dissociation energies.

A simple example of a first-principles approach is given by the B3LYP approximation
(and its materials counterpart, PBE0 \cite{ES99}).  The crucial part of a hybrid, which mixes
HF exchange with a GGA, is the fraction of exact exchange, $a$, which is about 20\%.
This was fixed once and for all in the definition of the functional, and all the thousands
of papers using B3LYP in Fig. \ref{papers} are gathering information on the same functional.
The amount of mixing can be rationally related to other calculations of atomization
energies \cite{PEB96}.
If authors {\em adjust} the amount of mixing to improve their results for
some system or property, this is {\em not} first principles.

Many of these tensions have been highlighted in recent years by the Minnesota functionals
developed by Truhlar and co-workers \cite{ZT08,ZT08b}.  These use the same basic ingredients
as the standard approximations, but optimize performance on a training set of energies by fitting up to
several dozen parameters.  They often produce more accurate results on systems close to those
trained on, but yield little insight into functional construction and development.
I consider these first-principles, because the functionals contain no parameters adjusted to the
system being calculated. But I also avoid altogether the e-word, because such functionals
illustrate the uselessness of that word in the context of DFT approximations.
\footnote{Interestingly, I have recently co-authored an approximation with about
10$^5$ `empirical' parameters \cite{SRHM11}. Don Truhlar, eat your heart out!}

\begin{figure}
\begin{center}
\includegraphics[width=3.5in]{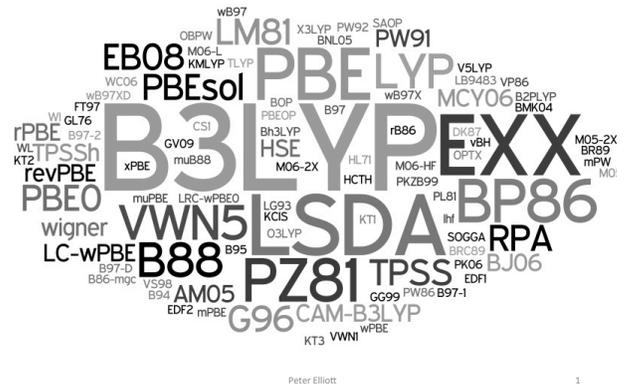}
\caption{The alphabet soup of approximate functionals available in a code near you. Thanks
to Peter Elliott for the artwork.}
\label{soup}
\end{center}
\end{figure}

Fig. \ref{soup} shows a random selection of some of the hundreds of XC approximations
that have been suggested over the years.  Many of these are in popular codes, some
of which even allow you to {\em design} your own functional.  
Clearly, when calculating a property to compare with experiment, one could keep 
trying functionals until agreement with the measured value is reached.
Not only is this contrary
to the entire spirit of DFT, 
it is certainly not first-principles, and is the worst form of empiricism.
The literature today is rife with such calculations, and the existence
of so many approximations, with so little guidance, is bringing the field into
disrepute.  Users should stick to the standard functionals (as most do, according
to Fig. \ref{papers}), or explain very
carefully why not.

\sec{Challenge: Digging deeper}
\label{dig}

As we have seen, the practice of modern DFT suffers a lack of detailed
understanding of how to approximate functionals.  We begin from local
approximations, and then create more accurate, sophisticated versions.
Based on insight and intuition, combining either exact conditions from
quantum mechanics or fitting specific systems, we make progress, but
only very slowly, and rarely without ambiguity.
A formally exact theory exists (Sec. \ref{exact}), but provides only
limited guidance about approximations.

I believe that a fundamental principle underlies the success of DFT, which
is that local approximations are a peculiar type of {\em semiclassical} approximation
to the many-electron problem \cite{CLEB10}.   For the last 6 years, with both my group and
many collaborators, I have been trying to uncover this connection, and make use
of it.  The underlying math is very challenging, and some must be invented.

Basic quantum textbooks have separate chapters on
perturbation theory and semiclassical approximations, but never relate the
two \cite{Griffiths}.   All modern many-body methods have their roots in treating the
interaction as a perturbation, since we find solving non-interacting
problems relatively easy.  But such treatments ignore the fact, proven by
Lieb more than a quarter of a century ago \cite{L81}, that TF theory (see Sec. \ref{hist})
becomes relatively exact for neutral atoms as $Z\to \infty$.  As detailed by
Schwinger \cite{S80} and others, this {\em is} the semiclassical limit \cite{ELCB08,PCSB06} mentioned above.
Dramatic confirmation of this fact is that TF also appears to yield the
exact ionization potential of atoms (averaged across a row) in this limit, and
that LDA-X recovers the HF result, including the oscillations across a row \cite{CSPB10}.
Most of our results so far have been confined to 1d systems \cite{ELCB08}, but
this insight lead to the restoration of the gradient expansion in PBEsol, which
cures many of the PBE problems with lattice parameters \cite{PRCV08}, and a
derivation of the parameter in the B88 functional \cite{EB09}.

\sec{Progress: Weak interactions}
\label{weak}
The ability to treat van der Waals is a recent (and ongoing) success
story for DFT.  In the 1990's, it was well-known that the standard
functionals could not yield correct long-range dispersion forces \cite{JG89}, i.e., 
their binding energy curves decay exponentially (with density overlap)
instead of $-C_6/R^6$, where $R$ is the separation and $C_6$ is the
van der Waals coefficient, determined from the the frequency-dependent
polarizabilities of the fragments \cite{GDP96}.   Because this excluded such a huge
number of important systems and properties (such as DNA, physisorption
on surfaces, most biochemistry, etc.), there were always adhoc methods
for adding back in dispersion, using pair potentials between atoms \cite{WY02}. 

Over about 20 years of research and many papers, the late David Langreth
with Bengt Lundqvist \cite{ALL96} 
and many other collaborators, developed an approximate
non-local ground-state density functional, call it LL, that has the right decay behavior and 
reasonably accurately captures these effects \cite{DRSL04}.  
LL is entirely non-empirical, using results from the uniform gas and
interactions between slabs of that gas to find such a form from first
principles.
While the initial implementation
of this functional was computationally expensive, a recent
algorithm of Soler \cite{SAGG02} 
made it much faster, so much so that its cost
is negligible beyond about 100 atoms.  This led to immediate implementation
in many codes worldwide, and there is now a plethora of calculations
with LL \cite{LKBA11}.

Simultaneous with this development, in quantum chemistry, Grimme \cite{G06} 
developed his DFT-D methodology that provides an empirical correction to DFT 
results in a systematic and accurate fashion.  The results for
small molecules in the S22 data set \cite{JSCH06} are extremely good.  DFT-D is much more accurate
for these systems than LL, but LL can be applied to all matter (except -possibly- metals),
including situations where pair-potentials cannot work.
A less empirical alternative to Grimme has been proposed by Tkatchenko 
and Scheffler \cite{TS09}, which produces
a scheme for calculating an additive correction for {\em any} functional and
has only slight empiricism.  This has recently been extended to include even
metals \cite{ZTPA11}. Over the next five years, one (or possibly two) of these schemes is
likely to become the standard method for including weak interactions in DFT.

\sec{Challenge: Time for a paradigm shift?}
\label{paradigm}
The KS equations, combined with the local density approximation,
were a reformulation of the electronic structure problem relative to TF theory.  By producing
a more demanding computational algorithm while lessening the fraction of
the total energy that needs approximating, a great leap forward in accuracy
and reliability was achieved.  But that was back in 1965.  Perhaps we are
at the end of that road in terms of useful approximations, and what is
needed now is a new paradigm which begins from different starting point.

\blue{Optimized effective potential (OEP):} \cite{SH53,TS76}
At some point, exact exchange (loosely, evaluating an orbital-dependent
functional in the KS scheme) seemed like a strong candidate, because
it allowed exchange to be evaluated exactly,  instead of being approximated.
This cures a multitude of
problems with local and semilocal approximations, such as better orbital energies,
avoiding self-interaction error, and better approximations
to the derivative discontinuity.  However, the technology for solving the
OEP equations efficiently has existed for at least a decade \cite{KLI92,G05,YW02}, but 
no general-purpose correlation approximation has been found that works
well with exact exchange in all situations.  Even worse, there are questions about how
well-defined the equations are in a basis \cite{SSD06}.

\blue{Random phase approximation (RPA):}
A more recent development \cite{EF11}, driven primarily by improvements in hardware
and algorithms, is the ability to solve the RPA equations efficiently
for systems up to about a hundred atoms.  These can now be
done {\em faster} than a conventional Hartree-Fock calculation.  Moreover,
bare RPA has many excellent features, including exact exchange and
a van der Waals contribution, but also has problems with atomization energies.
RPA is now implemented in both materials \cite{KF96} and chemical codes \cite{ABHH89}.  RPA makes
a promising candidate for a new baseline calculation, to which further
inexpensive approximations can be added, because it incorporates more
traditional theory (here, coupled-cluster \cite{SHS08})
at low computational cost.

\blue{Density matrix functional theory (DMFT):}
A darling of the chemistry community, and of increasing interest in
physics, is density-matrix functional theory (DMFT) \cite{DP78}, in which impressive
results have been gotten over the years.  The formalism is well-founded, 
using similar variational principles as DFT, due to Gilbert \cite{G75}, and 
a sequence of successive approximations and refinements have produced excellent
results at equilibrium bond lengths and for total energies of closed-shell
systems \cite{GPB05}.  Even gaps of insulators seem to come out well \cite{LSHD10}.
But the methods are not yet developed for general purpose,
as open shell systems and size-consistency remain issues.  If this methodology
ever does become popular, it would represent a true paradigm shift, as it
does not even use KS equations.  But, for this reason, it is difficult to
see how many of the impressive results of DFT approximations could be retained.

\sec{The future}

So, where does this leave us?   It is clearly both the best and worst of times for
DFT.  More calculations, both good and bad, are being performed than ever.
One of the most frequently asked questions of developers of traditional approaches to electronic
structure is: "When will DFT go away?".
Judging from Fig. \ref{papers}, no time soon.
Although based on exact theorems, as shown in Fig. \ref{vs}, these theorems
give no simple prescription for constructing approximations.  This leads to the
many frustrations of the now manifold users listed in Table \ref{frust}.  Without such
guidance, the swarm of available approximations of Fig. \ref{soup} will
continue to evolve and reproduce, ultimately undermining the entire field. 
Let us hope that some of the many excellent ideas being developed
by the community will come to fruition before that happens.

I am grateful for the support of NSF under grant no. CHE-1112442, and the assistance
of past and present students with the preparation of this perspective.

\bibliography{krefs}
\end{document}